\documentclass[aps,twocolumn,pre,showpacs]{revtex4}
\usepackage{graphics,amssymb,amsmath}

\begin{document}

\title{Performance of networks of artificial neurons: the role of clustering}

\author{Beom Jun \surname{Kim}}
\email{beomjun@ajou.ac.kr}
\affiliation{Department of Molecular Science
     and Technology, Ajou University, Suwon 442-749, Korea}

\begin{abstract}
The performance of the Hopfield neural network model is numerically
studied on various complex networks, such as the Watts-Strogatz network, 
the Barab{\'a}si-Albert network, and the neuronal network of
the C. elegans. Through the use of a systematic way of controlling the
clustering coefficient, with the degree of each neuron kept
unchanged, we find that the networks with the lower clustering 
exhibit much better performance.
The results are discussed in the practical viewpoint of
application, and the biological implications are also suggested.

\end{abstract}

%89.75.Hc Networks and genealogical trees
%84.35.+i Neural networks 
%87.18.-h Multicellular phenomena in biological system
%87.17.-d Cellular structure and processes

\maketitle

Recently, researches related with complex networks 
have been broadening territory beyond individual 
disciplines. Starting from pioneering works
of modeling complex networks~\cite{WS,BA}, the essential network
concepts have been successfully applied to 
various systems covering biological networks, food webs,
Internet, e-mail network, and so on~\cite{ref:network}.
In parallel to the studies of structural properties of 
complex networks, there have also been strong interest in
dynamic systems defined on networks. For examples, non-Hamiltonian
dynamic models such as epidemic spread~\cite{epidemic}, 
cascading failures~\cite{cascading}, 
synchronization~\cite{synch,smallmembrane}, the sandpile~\cite{sand}, and
the prisoner's dilemma game~\cite{pd}, as well as equilibrium
statistical physics models such as the Ising~\cite{ising} 
and the $XY$~\cite{smallXY} models have been investigated.

The network of neurons in biological organisms also takes the form of complex
networks~\cite{ref:network,realneuronnet,neuroanatomy,karbowski}.  While the  C.
elegans~\cite{WS,ref:network,cherniak} and the {\it in
vitro}~\cite{realneuronnet} neuronal networks have the high level of
clustering, the small characteristic path length, and the degree distribution
far from scale-free in common, the functional network of human
brains~\cite{dante} has recently been revealed to be scale-free.
Motivated by the neurons connected by synaptic couplings in biological
organisms, simple mathematical models of artificial neurons 
have been suggested~\cite{hopfieldreview}. One of the practical 
applications of such an artificial neural network can be found
in the Hopfield model~\cite{hopfieldreview}, which is used frequently 
in the pattern recognition. Very recently, there have been
studies of the Hopfield model of neurons put on the structure
of complex networks~\cite{neuronnet,costa}, with major focus on
how the topology, the degree distribution in particular,
of a network affects the computational performance of the
Hopfield model.   Also in the neuroscience, recent investigations
have revealed the close interrelationship between the brain activity and
the underlying neuroanatomy~\cite{neuroanatomy}.

\begin{table}[b]
\caption{Performance of Hopfield model on networks. All networks are
of the same size $N=280$ and the average degree $\langle k \rangle
\approx 14$. The neuronal network of the worm C. elegans, 
the Barab{\'a}si-Albert (BA) network, and the Watts-Strogatz (WS) 
networks at the rewiring probabilities $P=0.0, 0.1$, and 1.0  are used.  
Both the overlap $m$ and the input-output ratio $R$,  measuring
the performance of the network, increase 
as the clustering coefficient $\gamma$ is
decreased.  The maximum degree $k_{\rm max}$
is also shown to give some rough estimation of the broadness of the
degree distribution. }
\begin{ruledtabular}
\begin{tabular}{ccccc}
Network    & $k_{\rm max}$ &   $\gamma$  &  $m$  &  $R$ \\ \hline
WS($P=0.0$) & 14   &     0.69   &   0.689   & 0.603   \\
WS($P=0.1$) & 17   &    0.50   &   0.743  & 0.623   \\
C. elegans &  77 &   0.28   &   0.798   & 0.642  \\
BA       &   67 &  0.11   &   0.838   & 0.656  \\
WS($P=1.0$) & 22  &    0.05   &   0.881   & 0.672   \\
\end{tabular}
\end{ruledtabular}
\label{mytable}
\end{table}

Table~\ref{mytable} summarizes the clustering coefficients (see
Ref.~\onlinecite{WS} for the definition) and the performance
of the Hopfield model on various network structures~\cite{fullpaper} (see
below for details).  The difference in the performance,
measured by the overlap $m$ between the neuron state
and the stored pattern, has been previously attributed
to the distinct network topology, or  
the degree distribution more specifically~\cite{neuronnet}. Examining
Table~\ref{mytable} in more detail,  one can easily 
recognize the systematic dependence of the network performance 
on the clustering  property: As the clustering becomes
weaker, the performance is enhanced monotonically. The performance
detected by the ratio $R$ (the last column in Table.~\ref{mytable},
see Ref.~\onlinecite{forrest})
of the number of correctly recalled bits to  the  number of
synaptic couplings again shows the same behavior.

We in this work study the performance of the Hopfield
model on various network structures with focus on the
role of the clustering. We extend the
edge exchange method in Ref.~\onlinecite{sneppen} and 
suggest a systematic way to control the clustering coefficient
of a given network {\it without} changing the degree
of each vertex. A set of networks with the identical 
degree distribution but with various clustering coefficients
are generated and then used as the underlying network
structures for the numerical simulations of the Hopfield model.
Clearly revealed is that the computational
performance depends much more strongly on the 
clustering property than on the degree distribution.

\begin{figure}
\centering{\resizebox*{0.4\textwidth}{!}{\includegraphics{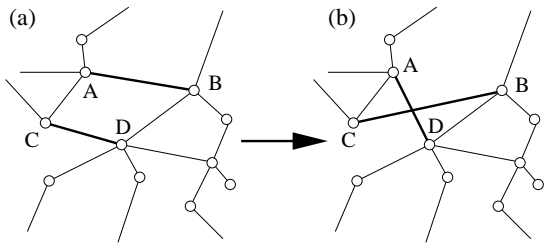}}}
\caption{Exchange of two edges: 
Two edges [A-B and C-D in (a)] are randomly picked and then 
rewired to have different end vertices [ A-D and C-B as in (b)].
The edge exchange keeps the degree of each vertex unchanged.
}
\label{fig:exchange}
\end{figure}

We begin with a brief review of the edge exchange method,
recently suggested by Maslov and Sneppen in Ref.~\onlinecite{sneppen}.
Two  edges, one connecting vertices A and B, and the other C and D,
are randomly chosen [Fig.~\ref{fig:exchange}(a)]. 
Each vertex changes its partner and the original edges 
A-B and C-D are altered to A-D and B-C  
as shown in Fig.~\ref{fig:exchange}(b). The important property
of the above edge exchange is that this process does not
change the degree of each vertex. A blind repetitions of
the above edge exchanges have been shown to destroy 
all degree-degree correlations~\cite{sneppen,dante}. 

Based on the above edge exchange method, we in the present paper
introduce an additional acceptance stage:
The edge exchange trial is accepted only when the new network 
configuration has higher (or lower)
clustering coefficient. This is identical to the standard
Monte Carlo (MC) simulation at zero temperature ($T$) with the Hamiltonian $H$ set to 
$ H = -\sum_v \gamma_v$ (or $H = \sum_v \gamma_v$), where $\gamma_v$ is the
clustering coefficient of the vertex $v$~\cite{WS}. In addition,
we also apply the standard MC scheme to the above Hamiltonian
at finite temperatures.
In either way, one can control
the clustering coefficient of a given network with the degree
of each vertex kept fixed. 

We use various network structures as underlying networks of the Hopfield
model. Since the real neuronal network of C. elegans has
$N=280$ vertices and the average degree $\langle k \rangle \approx  14$,
we generate various model networks of the size $N=280$ such as
Barab{\'a}si-Albert (BA) network with $M_0 = M = 7$ (starting
from $M_0$ vertices, one vertex with $M$ edges is added at each step
in the BA model~\cite{BA}), and the Watts-Strogatz (WS) networks with the
connection range $r=7$ at the rewiring probabilities $P=0$ (corresponding
to a regular local one-dimensional network), $P=0.1$ (well inside in
the small-world regime~\cite{WS}), and $P=1.0$ (corresponding
to a fully random network). All the networks have different topologies in
the sense that each has its own unique degree distribution.
After the original networks are  built, we decrease or increase
the clustering coefficients through the use of the above zero-temperature
MC method. Once the target value of the clustering coefficient
is reached, the current network structure is frozen and then used
as an underlying network structure for the Hopfield model.
We also perform MC at finite temperatures with the Hamiltonian
$H = -\sum_v \gamma_v$ to generate networks
at a given $T$ for comparisons.

In the Hopfield model of a neural network~\cite{hopfieldreview}, a neuron at the site $i$
can have two states, firing ($\sigma_i = 1$) or not firing ($\sigma_i = -1$).
The neuron state $\sigma_i(t)$ at time $t$ is determined from the configuration 
of other neurons at $t-1$ according to
\begin{equation} 
\label{eq:rule}
\sigma_i (t) = 
{\rm sgn}\left( \sum_j  \Lambda_{ij} W_{ij} \sigma_j(t-1) \right),
\end{equation} 
where $W_{ij}$ is the strength of the synaptic coupling between neurons
$i$ and $j$, and   
$\Lambda_{ij} (= 0, 1)$ describes the connection structure of the neural network.
For example, in the original version of the Hopfield neural network
the couplings are of the mean-field type and thus $\Lambda_{ij} = 1$ is
taken for any pair $(i,j)$ of two neurons. In the standard graph theory,
$\Lambda_{ij}$ is simply the $(ij)$-component of the so-called  
adjacency matrix, i.e., $\Lambda_{ij} = 1$ if two vertices $i$ and $j$
are connected while $\Lambda_{ij} = 0$ otherwise.
For simplicity, we in this work consider only undirected networks and 
accordingly $\Lambda$ is a symmetric matrix. Recently, Hopfield model
on the asymmetric directed BA network has been studied~\cite{costa}.
According to 
Eq.~(\ref{eq:rule}), the firing of neighbor neurons connected to the
$i$th neuron via excitatory synaptic  couplings ($W_{ij} > 0$) 
leads to the firing of $i$th neuron at next time step, while
inhibitory couplings ($W_{ij} < 0$) inhibits the firing of $i$th neuron.

For the task of the recognition of $p$ stored
patterns, the synaptic coupling strength $W_{ij}$ is usually given
by the Hebbian learning rule:
\begin{equation} 
W_{ij}  = \sum_{\mu=1}^p \xi_i^\mu\xi_j^\mu ,
\end{equation} 
where $\xi_i^\mu$ is the $i$th component of the $\mu$th stored
pattern vector and is quenched random variable taking values $1$ and $-1$ 
with equal probability. 
In this work, we present results only for $p=5$  (we also tried
$p=10$ and $20$ only to find insignificant qualitative differences).
The initial neuron state configuration $\{ \sigma_i (t=0) \}$
is produced from the one
of the pattern vectors, say the $\nu$th pattern, with 20\% error. In other words, 
$\sigma_i (0) = \xi_i^\nu$ for 80\% of the neurons,
while the other 20\% has the reversed bit $\sigma_i (0) = -\xi_i^\nu$.
As the neuron configuration evolves in time  by Eq.~(\ref{eq:rule}),
the overlap $m$ defined by
\begin{equation} 
m(t) \equiv \frac{1}{N} \sum_i \sigma_i(t) \xi_i^\nu
\end{equation} 
is measured as a function of time. The complete recognition of
the pattern $\nu$ gives $m=1$, which corresponds to the null Hamming distance 
$d(t) \equiv (1/N)\sum (\sigma_i(t)  - \xi_i^\nu)^2 = 2 - 2m(t)$.
We also measure the input-output ratio $R$ of the correctly recalled
pixels to the number of synaptic couplings, which is written as
$R = ( 1 + m ) p / \langle k \rangle $~\cite{forrest,fullpaper}.
The asynchronous dynamics, in which a neuron is chosen at random 
at each time step, is used throughout the present work.
After sufficiently long runs of dynamic evolution up to $t=2000$,
where the one time unit corresponds to the one whole sweep of all neurons, 
the last 200 time steps are used to make the time average of $m(t)$,
and disorder averages over 1000 different pattern realizations are performed.

\begin{figure}
\centering{\resizebox*{0.38\textwidth}{!}{\includegraphics{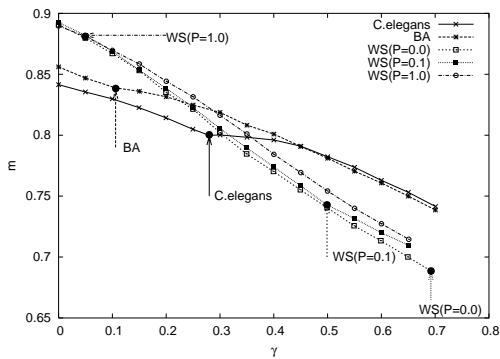}}}
\caption{Efficiency of the Hopfield neural network on the C. elegans
neuronal network, Barab{\'a}si-Albert (BA) network, 
and Watts-Strogatz (WS) network at the rewiring probabilities
$P=0, 0.1$, and $1.0$. The overlap $m$ between the stored pattern
and the neuron state measures the performance of each network
and is plotted as a function of the clustering coefficient 
$\gamma \equiv (\sum \gamma_v)/N $.
The big filled circles are for the original networks before
the edge exchange MC is applied (see text for details). 
It should be noted that at all points on each curve, not only the
degree distribution but also the actual degree of each neuron
is exactly identical.
Performance of each network monotonically decreases as the
clustering becomes stronger.
}
\label{fig:p5}
\end{figure}

\begin{figure}
\centering{\resizebox*{0.40\textwidth}{!}{\includegraphics{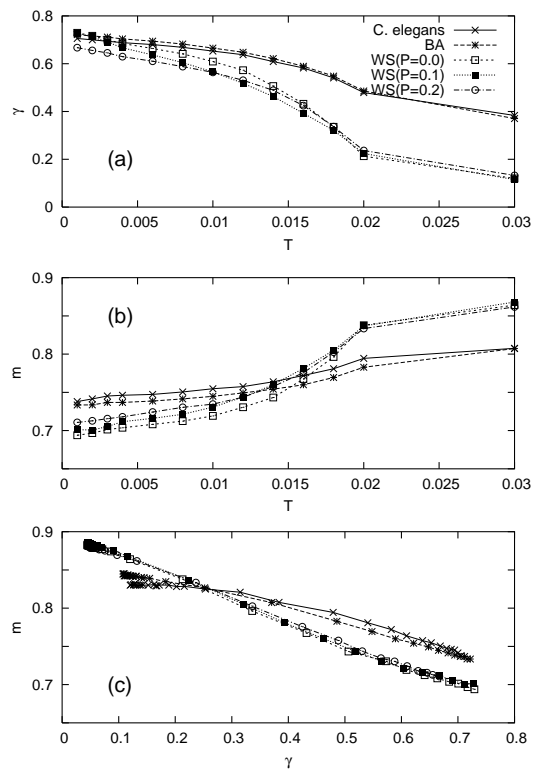}}}
\caption{Networks are generated from the finite-temperature edge-exchange MC 
for the Hamiltonian $H = -\sum_v \gamma_v$. (a) Clustering coefficient
$\gamma$ versus the temperature $T$. (b) The overlap $m$ versus $T$.
From (a) and (b), $m$ as a function of $\gamma$ is obtained in (c).
}
\label{fig:finiteT}
\end{figure}

Figure~\ref{fig:p5} is the main result of the present paper. For each
network, we start from the original network structure (denoted by the
big filled circle on each curve), and increase (or decrease) the
clustering coefficient by using the edge exchange zero-temperature MC
method described above. Once the target value of the clustering coefficient ($\gamma 
=0.0, 0.05, 0.10, 0.15, \cdots)$ is achieved the network structure is saved
and used for the Hopfield model simulations. We repeat this procedure
for 5-10 times to make an average over network structures.
We also present the result obtained for finite-temperature MC method
in Fig.~\ref{fig:finiteT}; We start from the original network at sufficiently
high temperature $T=1$ and use the standard Metropolis algorithm
applied for $H = -\sum_v \gamma_v$. At each $T$, we equilibrate the
network for $10^4$ MC steps and perform the Hopfield model dynamic
simulation. The temperature is then decreased slowly to make equilibration
at each $T$ more efficient.
It is to be noted that we have introduced
the finite temperature only when we generate networks by edge-exchange;
the time evolution of neuron states in Eq.~(\ref{eq:rule}) is still
deterministic. 

The most important conclusion one can make out of Figs.~\ref{fig:p5} and
\ref{fig:finiteT}(c) is that the performance of the Hopfield model on various
networks can be enhanced significantly if the clustering is weakened by the
simple method of edge exchanges without altering the degree of each neuron.
There are of course differences in performance among networks at the same
$\gamma$, e.g., for lower $\gamma$ WS networks are more efficient than the BA
and the C. elegans, while the trend becomes opposite for higher $\gamma$,
implying that the clustering coefficient is not the single parameter
controlling the performance.  However, Figs.~\ref{fig:p5} and
\ref{fig:finiteT}(c) strongly suggest that the efficiency of Hopfield neural
network depends much more strongly on $\gamma$ than on the degree distribution
as widely believed.  Consequently, we suggest that the difference in the
pattern recognition performance in Table~\ref{mytable} is strongly related with
the clustering property of each network. The result of the importance of the
clustering in the task of the pattern recognition by using the Hopfield model
has also some practical importance: It provides a novel way to enhance the
performance without adding more synaptic couplings~\cite{fullpaper}. For example, the simple 1D
Hopfield network can be made about 30\% more efficient by exchanging pairs of
edges.

The WS networks at $P=0.0, 0.1$, and 1.0 exhibit almost the same
performance for small clustering coefficient. As the clustering becomes
strong the WS network with $P=1.0$ has better performance than the other
smaller values of $P$~\cite{footWS}. However, the overall behavior does not show significant
differences.  On the other hand, it is interesting to note that
the BA and the C.elegans networks exhibit almost the same high 
performance at large $\gamma$, which appears to imply the importance of hub
vertices (the maximum degree in the C.elegans network is 77 while
it is 67 for the BA network as shown in Table~\ref{mytable}).

In the viewpoint of biological evolution, it is not clear
why the evolution chose the very structure of the neuronal
network of C.elegans. One expects that the actual detailed
structure of the neuronal network of C. elegans must have 
some advantage over other structures, which then leads
to a very crude expectation that the advantage may be detected
by the measurement of the performance of the Hopfield model,
as has been investigated in this work.
One somehow unlikely explanation is that the worm
is still in the evolutionary process of optimizing its
neuronal connection. The other explanation
can be the cost of the long-range synaptic couplings:
The actual connection topology of C.elegans can be the best
that the evolution can find if we consider the competition
between the performance  and the (energy) cost to make
long connections. This is somehow likely because the
present work does not take the geometric constraint
into account. 
Similar question on the competition between the
total axonal length (measuring energy cost for long-range connection)
and the characteristic path length (measuring the efficiency of
wiring structure) has been recently pursued in Ref.~\onlinecite{karbowski}
and the optimization of the neuronal-component placement on geometry 
has been investigated in Ref.~\onlinecite{cherniak}.
One can also think of other possibility that
the better performance by removing clustering has dark
side effect, like vulnerability of performance under
malfunctioning of neurons for example. In this respect,
it may also draw some interest that the performance curve
for C.elegans in Fig.~\ref{fig:p5} is the most flat one
around the point for the actual original network.

This work has been supported by the Korea Science
and Engineering Foundation through Grant 
No. R14-2002-062-01000-0 and Hwang-Pil-Sang research
fund in Ajou University. Numerical works have been
performed on the computer cluster Iceberg at Ajou University.


\begin{thebibliography}{99}
\bibitem{footWS} In the original WS network, the use of higher $P$
reduces the clustering coefficient~\cite{WS}. 
Consequently, both rewiring and edge-exchange can be used to enhance
the performance of Hopfield model on the WS network~\cite{fullpaper}.
\end{thebibliography}
\end{document}